\documentstyle[12pt,aasms4]{article}

\def\be{\begin{equation}}
\def\ee{\end{equation}}
\def\ba{\begin{eqnarray}}
\def\ea{\end{eqnarray}}

\def\la{\mathrel{\mathpalette\fun <}}
\def\ga{\mathrel{\mathpalette\fun >}}
\def\fun#1#2{\lower3.6pt\vbox{\baselineskip0pt\lineskip.9pt
        \ialign{$\mathsurround=0pt#1\hfill##\hfil$\crcr#2\crcr\sim\crcr}}}

\slugcomment{published in ApJ}

\begin{document}
\null\vspace{-62pt}
\begin{flushright}
Astrophys. J. 464, 114(1996)
\end{flushright}

\title{Statistics of Extreme Gravitational Lensing Events. I.\\
The Zero Shear Case}

\author{Yun Wang}
\affil{{\it NASA/Fermilab Astrophysics Center} \\
{\it Fermi National Accelerator Laboratory, Batavia, IL 60510-0500\\}
{\it email: ywang@fnas12.fnal.gov}}

\author{Edwin L. Turner}
\affil{{\it Princeton University Observatory} \\
{\it Peyton Hall, Princeton, NJ 08544\\}
{\it email: elt@astro.princeton.edu}}

\vspace{.4in}
\centerline{\bf Abstract}
\begin{quotation}
For a given source and lens pair, there is a thin on-axis tube-like
volume behind the lens in which the radiation flux from the source 
is greatly increased due 
to gravitational lensing.
Any objects (such as dust grains) which pass through such a thin
tube will experience strong bursts of radiation, i.e., 
Extreme Gravitational Lensing Events (EGLE). 
We study the physics and statistics of EGLE for the case in which finite
source size is more important than shear.
One of the several possible significant astrophysical effects is
investigated with an illustrative calculation.
\end{quotation}


\section{Introduction}
Gravitational lensing has been studied almost exclusively in the context
of direct observation of lensed sources from the Earth or a spacecraft
(\cite{BNrev92}, \cite{Book92}).
However, every astrophysical
object receives the
light from sources lensed by intervening massive objects.
The most powerful gravitational lensing events occur when the source, lens,
and a target object are nearly on-axis. Such events are extreme in magnification, 
and rare in occurrence for any given target.  By considering arbitrarily
located targets, we can study the statistics of Extreme Gravitational 
Lensing Events (EGLE). EGLE can have a significant effect on certain 
objects,
especially fragile components of the interstellar medium, 
such as molecules or dust grains.

For a pair of source and lens, an EGLE occurs when a moving target object
crosses the source-lens line behind the lens.
The maximum magnification of the source seen by the target can be 
extremely large, limited only by the source size and the shear on the lens.
As the target moves away from the line connecting the source and the lens, 
the magnification of the source decreases. The duration of an EGLE depends 
on the velocity of the target and the size of the high flux region.  
A slowly moving target in the neighborhood
of a pair of small-size source and slightly-sheared lens can experience
a strong burst of radiation due to the lensing of the source, which
may be sufficient to affect the target's properties.

If the target moves a distance $d$ away from the line 
connecting the source and the lens, it is equivalent to the source 
moving an angular distance of $y$ from the optical axis (the line connecting 
the lens and the target). Measuring $y$ in units of the angular 
Einstein radius, we have
\be
\label{eq:O-S}
y \simeq \left( \frac{D_{\rm ds}}{D_{\rm d}} \right)\,
\frac{d}{D_{\rm s} \theta_{\rm E}},
\ee
where $D_{\rm ds}$, $D_{\rm s}$, and $D_{\rm d}$ are angular 
diameter distances between the lens and source, target and source, 
target and lens respectively. $\theta_{\rm E}= \sqrt{ 2 R_{\rm S} 
D_{\rm ds}/(D_{\rm d} D_{\rm s})}$ is the angular Einstein radius.  
$R_{\rm S}=2GM$ is the Schwarzschild radius of the lens with mass $M$. 
Eq.(\ref{eq:O-S}) is 
the small angle approximation, valid for $d \ll D_{\rm d}$.

We can write $\theta_{\rm E}$ as
\be
\theta_{\rm E}= 10^{-6} \times \sqrt{ \left( \frac{M}{5\times 10^6 M_{\odot}}
\right)\, \left( \frac{1 \,{\rm Mpc}}{D_{\rm d}}\right)\,
\frac{D_{\rm ds}}{D_{\rm s}} }.
\ee
The dimensionless radius of a source with physical radius $\rho$ is
defined as
\be
R \equiv \frac{\rho}{D_{\rm s} \theta_{\rm E}}
= \left(\frac{\rho}{1 {\rm pc}}\right)\,
\left( \frac{10^{-6}}{ \theta_{\rm E}} \right) \,
\left(\frac{1 \,{\rm Mpc}}{D_{\rm s}}\right).
\ee

For a given pair of lens and source, the shear $\gamma$ on the lens
due to other lensing objects near the line-of-sight 
is the same order of magnitude as the 
optical depth for microlensing, $\tau$, the probability that the 
source is lensed. We find
\be
\label{eq:gamma}
\gamma = \sqrt{2}\,\, \zeta\left(\frac{3}{2}\right)\, \tau
\simeq 3.7\, \tau.
\ee
$\zeta\left(x\right)$ is the Riemann zeta function.
Not surprisingly, the statistics of EGLE is much more complicated for 
sheared lenses than for isolated lenses ($\gamma=0$). 
Since $\tau= \Omega_{\rm L} z_{\rm Q}^2/4$
($\Omega_{\rm L}$ is the critical density fraction in lenses and $z_{\rm Q}$
is the source redshift), the shear $\gamma$ is probably
small in the low redshift Universe. (\cite{Turner80,Turner84})

The lens model for a perturbed Schwarzschild lens (i.e., point
mass lens with shear) has been investigated by 
Chang and Refsdal (1979, 1984) and Subramanian and Chitre (1985).
For $\gamma\ll 1$, the caustic is an astroid shaped curve with four
cusps. The magnification of extended sources by a perturbed Schwarzschild lens
has been studied by Chang (1984), Schneider and Weiss (1987).
For a finite source with dimensionless radius $R$ which crosses 
the optical axis, Figure 1 shows the typical lightcurves for $\gamma=0$ 
(solid line), $\gamma=R/2$ (dotted line), $\gamma=R$ (short dashed line), 
$\gamma=5R$ (long dashed line), and $\gamma=10R$ (dot-dashed line) 
respectively; Figure 2 shows the corresponding cross-sections of magnification. 
Clearly, shear is not important for $\gamma <R \ll 1$. 
Note that the peaks in the lightcurves in Fig.1 correspond to caustic
crossing.

In this paper, we study the statistics of EGLE for sources with 
small dimensionless radius $R$ and isolated lenses ($\gamma=0$). 
We generally follow the notation and conventions of Schneider et al. (1992).
We will discuss the statistics of EGLE for lenses with small shear 
$\gamma$ elsewhere (\cite{Turner96}).

\section{Basic statistics}

In this section, we discuss mean and rms magnifications, as well
as integrated excess flux (IEF) seen by a target,
for finite sources and Schwarzschild lenses (i.e., point mass lens, 
no shear).

The magnification of a finite source by a Schwarzschild lens has been
studied by Bontz (1979) and Schneider (1987).
The magnification of a source with dimensionless radius $R$ is given by
(\cite{SC87.4})
\be
\label{eq:mu(R)}
\mu_{\rm e} (y, R)  \simeq 
\left\{ \begin{array}{ll}
\frac{1}{R} \,\zeta\left(\frac{y}{R}\right) & \mbox{for $ y \la 5R$}\\
\,\,\\
\mu_{\rm p}(y)= \frac{(y^2+2)}{y \sqrt{y^2+4}}
& \mbox{for $ y \ga 5R$} 
\end{array}
\right.
\ee
where
\be
\zeta(w) = \frac{2}{\pi} \int^1_0 {\rm d} x\, x \int^{\pi}_0
\frac{ {\rm d}\phi} { \sqrt{w^2+x^2+2wx \cos\phi}}.
\ee
$\zeta(0)=2$, $\zeta(w>0)$ can be easily integrated numerically.

Using Eq.(\ref{eq:O-S}), we can define the half-widths of
observables in the source plane.
Let us define the half-width of the light curve seen by the target
to be $y^{\rm w}_{\rm HM}$, the source's distance to the optical axis
when its magnification goes down to $1/2$ its maximum $\mu_{\rm max}$.
It's straightforward to find:
\be
\mu_{\rm max} = \frac{2}{R}, \hskip 1cm 
y^{\rm w}_{\rm HM} \equiv y(\mu=0.5\,\mu_{\rm max}) \simeq 1.145 R.
\ee
The mean, root mean square, and root variance magnifications are
given by
\ba
&&\langle \mu \rangle \left(y\leq y^{\rm w}_{\rm HM}\right)
\simeq \frac{1.5583}{R}, \hskip 1cm \sqrt{\langle \mu^2 \rangle}\,
\left(y\leq y^{\rm w}_{\rm HM}\right)
\simeq \frac{1.588}{R}, \nonumber\\
&&\sqrt{\langle (\mu- \langle \mu \rangle)^2 \rangle }\,
\left(y\leq y^{\rm w}_{\rm HM}\right) \simeq \frac{0.3}{R}.
\ea

We are also interested in the integrated excess flux $F$ seen by a moving 
target. Let us define $t=0$ to be the moment when the moving target 
crosses the line connecting the source and lens. For a target moving 
at constant velocity $v$, its distance from the line connecting the 
source and lens is $d=vt$. Using Eq.(\ref{eq:O-S}), we have
\ba
F(t) &\equiv& \int_0^t {\rm d}t\, \left[\mu_{\rm e}(t, R)-1\right]
\nonumber \\ 
&=&\frac{D_{\rm d}}{ D_{\rm ds}} \,\frac{D_{\rm s} \theta_{\rm E}}{v} 
\int^y_0 {\rm d}y\, \left[\mu_{\rm e}(y, R)-1\right] \equiv 
\frac{D_{\rm d}}{ D_{\rm ds}} \,\frac{D_{\rm s} \theta_{\rm E}}{v} \, 
\overline{F}(y).
\ea
Using Eqs.(\ref{eq:mu(R)}), we find
\be
\overline{F}_{\rm total} \equiv \int^\infty_0 {\rm d}y\, 
\left[\mu_{\rm e}(y, R)-1 \right] \simeq 1.27 -\ln R,
\ee
for $R \la 0.05$. We define the half-width of the integrated excess flux
(IEF) to be $y^{\rm w}_{\rm IEF}$, the source's distance from the
optical axis when the IEF seen by the target is half the total IEF, i.e.,
$\overline{F}(y^{\rm w}_{\rm IEF})=\overline{F}_{\rm total}/2$. We find
\be
y^{\rm w}_{\rm IEF}(R) \simeq 0.287 \sqrt{R}.
\ee
The corresponding mean, root mean square, and root variance magnifications 
are 
\ba
&&\langle \mu \rangle \left(y\leq y^{\rm w}_{\rm IEF}\right) 
\simeq \frac{6.97}{\sqrt{R}}, \hskip 1cm \sqrt{\langle \mu^2 \rangle}\,
\left(y\leq y^{\rm w}_{\rm IEF}\right) \simeq \sqrt{ \frac{24.28}{R}\, 
\ln\left(\frac{1.466}{\sqrt{R}}\right)},  \nonumber\\
&&\sqrt{\langle (\mu- \langle \mu \rangle)^2 \rangle }\,
\left(y\leq y^{\rm w}_{\rm IEF}\right) \simeq \frac{4.93}{\sqrt{R}} 
\,\sqrt{ \ln\left(\frac{0.2}{\sqrt{R}} \right) }.
\ea

It is useful to consider only the $y<1$ regime, the half-width
of microlensing events.  The corresponding mean, 
root mean square, and root variance magnifications are 
\ba
&&\langle \mu \rangle\left(y\leq 1\right) \simeq 2.236-0.06\,R, \hskip 1cm
\sqrt{\langle \mu^2 \rangle}\,\left(y\leq 1\right) \simeq \sqrt{ 2\, 
\ln\left(\frac{7.531}{R}\right)}, 
\nonumber\\
&&\sqrt{\langle (\mu- \langle \mu \rangle)^2 \rangle }\,\left(y\leq 1\right) 
\simeq \sqrt{ 2\, \ln\left( \frac{0.6182}{R} \right) }.
\ea

Generally, for $y \ga 5R$, we have
\ba
&&\langle \mu \rangle\left(\leq y\right) \simeq \frac{\sqrt{y^2+4}}{y}
-\frac{0.06R}{y^2}, \nonumber\\
&&\langle \mu^2 \rangle\,\left(\leq y\right) \simeq 1+  \frac{2}{y^2}\left[ 
\ln\left( \frac{10.2135}{R} \right)- \ln\left( \frac{\sqrt{y^2+4}}{y}
 \right) \right].
\ea
For $y \rightarrow \infty$, $\langle \mu \rangle$=1,
as required by flux conservation, and $\langle \mu^2 \rangle$=1.
However, it is not physical to average over $y$ from $0$ to infinity.
For a given source, there is a natural cut-off $y_{\rm max}$ 
which is given by the source's distance to the nearest lens, 
i.e., $y_{\rm max} \sim 1/\sqrt{\tau}$, where $\tau$ is the optical depth.

\section{EGLE volume statistics for a point source}

Let us consider a point source S with luminosity $L_{\rm S}$, 
being lensed by a lens L with Schwarzschild radius $R_{\rm S}$
(mass $M$) at a distance $D_{\rm ds}$.
Let $Q$ denote either the magnification or the flux of the source 
seen by the target.
In a narrow tube-shaped volume $V_{\rm SL}$ behind the lens, 
which extends from the lens and tapers off to infinity, 
$Q$ exceeds some value $q$. The cross-section of the tube is 
\be
\label{eq:x-sec}
\sigma(q) = \pi d^2=\pi\left[\left(\frac{D_{\rm d}}{D_{\rm ds}}\right)^2\, 
D_{\rm s}^2\theta^2_{\rm E}\right]\, y^2(q),
\ee
where we have used Eq.(\ref{eq:O-S}). $y(q)$ is the source's dimensionless
distance from the optical axis when $Q$ equals $q$. Hence
\be
\label{eq:V_SL}
V_{\rm SL}(q)= \int^{D_{\rm d}(q)}_0 {\rm d}D_{\rm d}\, \sigma(q).
\ee
Summing over S gives the total volume $V_{\rm L}$ in which $Q$
exceeds $q$ for a given lens L; further summing over 
L gives the total volume $V_{\rm tot}(>q)$.
We use $D_{\rm s}= D_{\rm d}+D_{\rm ds}$ for simplicity in our calculations.

Let us consider the volume $V_{\rm SL}(f)$ behind the lens
in which the flux from the source exceeds $f$.
In the absence of magnification, the flux from the source is
$f_0 = L_{\rm S}/(4 \pi D_{\rm s}^2)$. The magnified flux $f = \mu f_0$. 
Since we are only interested in high magnification events, we use 
$y(f) \simeq 1/\mu=f_0/f$ in calculating the cross-section $\sigma(f)$ of
$V_{\rm SL}(f)$. We find
\be
\label{eq:xectp}
\sigma(f, D_{\rm d}) = \frac{2\pi R_{\rm S}}{D_{\rm ds}}\cdot \frac{D_{\rm d}}
{D_{\rm s}^3} \left( \frac{L_{\rm S}}{4\pi f}\right)^2.
\ee
$\sigma(f, D_{\rm d})$ is maximum at $D_{\rm d}=D_{\rm ds}/2$. We find
\be
\label{eq:xectpmax}
\sigma^{\rm max}_{\rm p}(f) = \frac{8\pi R_{\rm S}}{27 D_{\rm ds}^3}
 \left( \frac{L_{\rm S}}{4\pi f}\right)^2.
\ee
We can define $\sigma^{\rm max}_{\rm p}(f)$ to be the characteristic 
cross-section of the high-flux ($>f$) tube. The lens which is closest to the
source has the thickest high-flux tube behind it.

Using Eq.(\ref{eq:xectp}), we obtain
\be
\label{eq:ps V(f)}
V_{\rm SL}(f) \simeq \frac{\pi R_{\rm S}}{D_{\rm ds}^2}
\left( \frac{L_{\rm S}}{4\pi f}\right)^2, \hskip 1cm
V_{\rm L}(f) = 4\pi^2 n_{\rm S} R_{\rm S} D_{\rm c}
\left( \frac{L_{\rm S}}{4\pi f}\right)^2,
\ee
where $n_{\rm S}$ is the number density of sources, and $D_{\rm c}$
is the size of the system. 
In realistic astrophysical systems, the spatial distribution of sources,
lenses and targets is complex and varied.  In this paper we will
consider all systems to be spheres of diameter $D_{\rm c}$ with all three
of these populations distributed uniformly and randomly throughout, a
simple but reasonably general approximation.
In effect, $D_{\rm c}$ is the maximum distance between
a lens and a source.

Let ${\cal F}_{\rm L}(f)$ be the volume fraction of space in which the 
flux from the source exceeds $f$ due to gravitational lensing.
${\cal F}_{\rm L}(f)$ should be compared with the volume fraction of space 
${\cal F}_{\rm S}(f)$ in which the flux from the source exceeds $f$ 
due to being close to the source. We have
\ba
&&{\cal F}_{\rm L}(f) = n_{\rm L}  V_{\rm L}(f)=
\frac{3}{2}\, \tau N_{\rm S} \left(\frac{f}{f_{\rm min}}\right)^{-2},
\hskip 1cm {\cal F}_{\rm S}(f)= N_{\rm S} 
\left(\frac{f}{f_{\rm min}}\right)^{-3/2}, \nonumber\\
&&\frac{{\cal F}_{\rm L}(f)}{{\cal F}_{\rm S}(f)}
= \frac{3}{2}\, \tau \left(\frac{f}{f_{\rm min}}\right)^{-1/2},
\ea
where $\tau$ is the optical depth, $N_{\rm S}$ is the total
number of sources, and $f_{\rm min}=L_{\rm S}/(4\pi D_{\rm c}^2)$.
Note that the volume weighted rms flux due to lensing diverges
logarithmically.

The average flux from the 
general population of sources is $n_{\rm S} D_{\rm c} L_{\rm S}$.
Let us define relative flux
\be
f'= \frac{f}{n_{\rm S} D_{\rm c} L_{\rm S}}.
\ee
Let ${\cal F}_{\rm L}(f')$ and ${\cal F}_{\rm S}(f')$ be the volume fractions
of space in which the relative flux from the source exceeds $f'$ due to lensing
and due to being close to the source respectively.
We have
\ba
&&{\cal F}_{\rm L}(f', \rho=0) = n_{\rm L}  V_{\rm L}(f', \rho=0)=
\frac{\tau}{ 6 N_{\rm S} f'^2},
\nonumber\\
&&\frac{{\cal F}_{\rm L}(f', \rho=0)}{{\cal F}_{\rm S}(f')}
= \frac{\tau}{2}\, \left(\frac{3}{N_{\rm S}}\right)^{1/2}f'^{-1/2}.
\ea

\section{EGLE volume statistics for a finite source}

Now let us consider a source S with physical radius $\rho$ and luminosity 
$L_{\rm S}$, being lensed by a lens L with Schwarzschild radius $R_{\rm S}$
(mass $M$) at a distance $D_{\rm ds}$.
The tube-shaped volume $V_{\rm SL}(f,\rho)$ behind the lens
in which the flux from the source exceeds $f$ has finite length
$D^{\rm m}_{\rm d}(f, \rho)$, because of the finite size of the source.

For a finite source with dimensionless radius $R$, $\mu_{\rm max}=2/R$. 
Let us define a parameter $\alpha(f)$ which measures the maximum magnification
of the source relative to the flux $f$,
\begin{equation}
\label{eq:alpha(f)}
\alpha \equiv \frac{8R_{\rm S} D_{\rm c}}{\rho^2} \left(\frac{L_{\rm S}}
{4\pi D_{\rm c}^2 f}\right)^2. 
\end{equation}

The tube volume $V_{\rm L}(f, \rho)$ has the cross-section 
$\sigma(f, \rho, D_{\rm d})$ which vanishes at $D_{\rm d}=0$, 
$D^{\rm m}_{\rm d}$. To calculate the cross-section $\sigma(f, \rho)$, 
we need to know $y(f)$ [see Eq.(\ref{eq:x-sec})], which can be found by
inverting $\mu(y)= f/f_0$ numerically. $\sigma(f, \rho)$ can be written as
\begin{equation}
\sigma(f, \rho, D_{\rm d})  =\pi \rho^2 \times d^2/\rho^2
=\pi \rho^2 \, \overline\sigma(\alpha, D_{\rm d}),
\end{equation}
for given $D_{\rm ds}$. 
Figure 3 shows the cross-section $\sigma(f, \rho, D_{\rm d})$ with
$\alpha(f)=4$, for $D_{\rm ds}=0.2\,D_{\rm c}$ (solid line), $0.5\,D_{\rm c}$
(long dashed line). The lens which is closest to the source has the thickest 
tube of high flux behind it, as in the point source case.

To simplify the calculation for the volume fractions, 
let us approximate Eq.(\ref{eq:mu(R)}) with
\be
\label{eq:app mu_e}
\mu_{\rm e}(y, R) \simeq \left\{ \begin{array}{ll}
\mu_{\rm p}(y) & \mbox{for $\mu < \mu_{\rm max}$}\\
\,\,\\
\mu_{\rm max} & \mbox{elsewhere} 
\end{array}
\right.
\ee
where $\mu_{\rm p}(y)$ is the point source magnification,
and $\mu_{\rm max}=2/R$.
Eq.(\ref{eq:app mu_e}) is reasonably good for $\mu \la 1/R$. 
Figure 3 shows the approximate cross-sections obtained by using 
Eq.(\ref{eq:app mu_e}) for $D_{\rm ds}=0.2\,D_{\rm c}$ (short dashed line), 
$0.5\,D_{\rm c}$ (dot-dashed line). The dotted lines indicate
$D^{\rm m}_{\rm d}$. This approximate cross-section always 
{\it under-estimates} the true cross-section; the difference increases with
decreasing $\alpha(f)$ (large $\rho$ or $f$), but it is negligible for
our purposes.

Now let us derive the length of the tube-volume 
$V_{\rm SL}(f,\rho)$. Note that $f =\mu f_0 \leq \mu_{\rm max} f_0$. Let 
$f =\mu_{\rm max} f_0$ at $D_{\rm d}(f)=D_{\rm d}^{\rm m}(f)$,
i.e., the flux is equal to $f$ on the line SL connecting the source
and lens. For given $D_{\rm d}$, the flux decreases away from line SL, 
hence the volume in which the flux exceeds $f$ converges to a point 
at $D_{\rm d}(f)=D_{\rm d}^{\rm m}(f)$.
For $D_{\rm d} > D_{\rm d}^{\rm m}$, the volume in which
the flux exceeds $f$ is zero. For a given pair of source and lens,
$D_{\rm d}^{\rm m}(f)$ gives the length of the tube volume in
which the flux exceeds $f$. To find $D_{\rm d}^{\rm m}(f)$, we write
$f =\mu_{\rm max} f_0$ as
\be
\frac{D_{\rm d}^{\rm m}}{D_{\rm c}} \left(\frac{D_{\rm d}^{\rm m}}{D_{\rm c}} 
+\frac{D_{\rm ds}}{D_{\rm c}} \right)^3= \frac{8R_{\rm S} D_{\rm ds}}
{\rho^2} \left(\frac{L_{\rm S}}{4\pi D_{\rm c}^2 f}\right)^2 =\alpha
\, \left(\frac{D_{\rm ds}}{D_{\rm c}} \right).
\ee
The above equation can be solved analytically for $D_{\rm d}^{\rm m}(f)$. 
Let us define
\be
x \equiv \frac{D_{\rm ds}}{D_{\rm c}},  
\hskip 2cm \omega \equiv \frac{x}{\alpha^{1/3}}.
\ee
We find
\be
\label{eq:D_d(f)}
\frac{D_{\rm d}^{\rm m}(f)}{D_{\rm ds}} = \frac{1}{2} \left\{
\frac{\left[\omega \, u(\omega)\right]^{3/2}}{2} +
\sqrt{ \frac{1}{2} \left[1+ \left[\omega \, u(\omega)\right]^{-3/2}\right]
+u(\omega)}-\frac{3}{2} \right\}\equiv g(\omega), 
\ee
where
\be
u(\omega)= \frac{3}{4} \left[b(\omega)+\frac{1}{b(\omega)}+1\right]^{-1},
\hskip 1cm b(\omega)= \frac{3\omega}{4} \left[\frac{1}{2}+
\sqrt{ \frac{1}{4} +\left(\frac{4}{3\omega}\right)^3} \,\right]^{2/3}.
\ee
$g(\omega)= D_{\rm d}^{\rm m}(f)/D_{\rm ds}$ is shown in Figure 4. 
Given the separation between the source and the lens, 
$x=D_{\rm ds}/D_{\rm c}$, the length of the tube volume behind the lens
in which the flux exceeds $f$ is given by $D_{\rm d}^{\rm m}(f)/D_{\rm c}
= x \,g(\omega)$, where $\omega =x\alpha^{-1/3}$.
The tube length is of order $D_{\rm c}$ for $\alpha$ of order 1.

The function $g(\omega)$ has the following asymptotic behavior:
\be
g(\omega) = \left\{ \begin{array}{ll}
\omega^{-3} & \mbox{for $\omega \gg 1$}\\
\omega^{-3/4} & \mbox{for $\omega \ll 1$} 
\end{array}
\right.
\ee
Thus we have
\be
\frac{D_{\rm d}^{\rm m}(f)}{D_{\rm c}} = \left\{ \begin{array}{ll}
\alpha(f)/x^2 & \mbox{for $\alpha^{1/3} \ll x$}\\
\left[\alpha(f)\,x\right]^{1/4} & \mbox{for $\alpha^{1/3} \gg x$} 
\end{array}
\right.
\ee

Substitution of Eq.(\ref{eq:D_d(f)}) into Eq.(\ref{eq:V_SL}) gives 
$V_{\rm SL}(f,\rho)$. Using Eq.(\ref{eq:app mu_e}) and
$\mu_{\rm p}(y) \simeq 1/y$ (for high magnification events) in
computing $\sigma(f, \rho)$, we find
\be
V_{\rm L}(f, \rho)= 4\pi^2 n_{\rm S} R_{\rm S} D_{\rm c}
\left(\frac{L_{\rm S}}{4\pi f}\right)^2 \, I(\alpha),
\ee
where
\be
\label{eq:I(alpha)}
I(\alpha) = \alpha^{1/3}\, \int^{\alpha^{-1/3}}_0 \frac{ {\rm d}\omega}
{\left[1+ 1/g(\omega)\right]^2}.
\ee
For high flux, $\alpha(f) \ll 1$, $I(\alpha) = 0.3583\, \alpha^{1/3}$.
Note that $I(\alpha)= V_{\rm L}(f, \rho)/V_{\rm L}(f, \rho=0)$.
In the point source limit, $\alpha \gg 1$, $I(\alpha)=1$.
We show $I(\alpha)$ in Figure 5.

Given $I(\alpha)= A \alpha^{\beta}$, with 
${\cal F}_{\rm L}(f)$ and ${\cal F}_{\rm S}(f)$ denoting the volume fractions
of space in which the flux from the source exceeds $f$ due to lensing
and due to being close to the source respectively, we find
\ba
&&{\cal F}_{\rm L}(f, \rho) = \frac{A}{4\cdot 2^{\beta}\pi^{2\beta}}
\, n_{\rm S} n_{\rm L} R_{\rm S}^{1+\beta} L_{\rm S}^{2+2\beta}
f^{-(2+2\beta)} D_{\rm c}^{1-3\beta} \rho^{-2\beta},
\nonumber\\
&&\frac{{\cal F}_{\rm L}(f, \rho)}{{\cal F}_{\rm S}(f)}
= \frac{3\pi^{1/2-2\beta} A}{2^{1+\beta}}\, n_{\rm L} 
R_{\rm S}^{1+\beta} L_{\rm S}^{1/2+2\beta}
f^{-(1/2+2\beta)} D_{\rm c}^{1-3\beta} \rho^{-2\beta}.
\ea

\section{Possible astrophysical effects}

For a population of sources (with number density $n_{\rm S}$)
lensed by a population of lenses (with number density $n_{\rm L}$),
the physical picture for EGLE is a complex network of thin high-flux
tubes, at each knot sits a lens, and each tube line points away from a source.
In other words, a given lens has one high-flux tube coming out of it because 
of each source, and a given source induces one high-flux tube behind each lens.
 
To roughly survey possible astrophysical effects of EGLE, we construct
tables of possible source and lens populations.
Table 1 lists a few types of small sources with high luminosity. 
Note that the space density $n_{\rm S}$ associated with transient sources such
as $\gamma$-ray bursts and supernovae includes the finite lifetime factor;
in other words, it is the density of sources shining at a particular moment.
Table 2 gives two possible lens populations. 
Again, we have assumed that the sources and the lenses are mixed and both
are distributed uniformly in a volume with size $D_{\rm c}$. Here we only
consider the volume fractions of space occupied by the high-flux EGLE tubes
(which are determined by the sources and lenses only); the results
derived in this Section apply to arbitrary targets uniformly distributed
in the same volume with the sources and the lenses.

In the context of EGLE, the relevant dimensional physical quantities are:
lens mass $M$, size of the lens-source distribution $D_{\rm c}$,
source size $\rho$, source luminosity $L_{\rm S}$, minimum lensed flux 
from the source $f$. All these collapse into a single dimensionless parameter
$\alpha(f)$ [see Eq.(\ref{eq:alpha(f)})], which
measures the maximum magnification of the source relative 
to the flux $f$. In astrophysical units, we write
\begin{eqnarray}
\log(\alpha) &=& -5.62+0.8\,m_{\rm bol}+ 2\,\log\left( 
\frac{L_{\rm S}}{L_{\odot}}\right)+ \log\left( \frac{M}
{M_{\odot}}\right)- 2\, \log\left( \frac{\rho}{R_{\odot}}\right)
\nonumber\\
&&\hskip 2cm -3\, \log\left( \frac{D_{\rm c}}{1\,{\rm kpc}}\right)
\label{eq:alphaAstro}
\end{eqnarray}
where the minimum flux $f$ is measured by $m_{\rm bol}$. 
Table 1 shows $\log(\alpha)-0.8 \,m_{\rm bol}$ for the listed sources 
with $M=M_{\odot}$, i.e., for lensing by stars.

Let us first consider the physical dimensions of a high-flux tube 
behind a lens, in which the flux exceeds $f$.
Using Eq.(\ref{eq:D_d(f)}), we find
\be
\log\left( \frac{D_{\rm d}^{\rm m}} {\rm cm} \right)=
21.49 +\log\left( \frac{D_{\rm ds}}{\rm kpc}\right) +\log g(\omega),
\ee
where $\omega=(D_{\rm ds}/D_{\rm c}) \alpha^{-1/3}$.
$g(\omega)$ is given by Eq.(\ref{eq:D_d(f)}) and 
shown in Fig.4. For $\alpha \ll 1$, $\omega \gg 1$ at a
given $D_{\rm ds}$, $g(\omega) \simeq \omega^{-3}$; i.e., the length
of the high-flux tube decreases sharply for small $\alpha$. In Fig.6,
we show $D_{\rm d}^{\rm m}$ versus $m_{\rm bol}$  
for lensing by stars, the sources are $\gamma$-ray bursts (solid line),
QSO (X-ray) (dotted line), QSO (UV-opt) (short dashed line), 
Galactic supernovae (long dashed line), neutron stars (dot-short dashed line),
hot O stars (dot-long dashed line), and hot B stars 
(short dash-long dashed line) respectively. 
Note that we have taken $\gamma$-ray bursts to be cosmological in origin
(\cite{gamma-rayBurst}); models which attribute them to local sources
would predict smaller EGLE effects.

Since the cross-section of the high-flux tube for a finite source 
does not deviate significantly from that of a point source [see
Fig.3], we define the characteristic radius $a$ of the high-flux 
tube to correspond to the maximum cross-section (along the optical axis)
of the high-flux tube for a point source [see Eq.(\ref{eq:xectpmax})]. 
We find
\be
\log\left( \frac{a} {\rm cm} \right)= 7.38 +0.4\, m_{\rm bol}
+\log\left(\frac{L_{\rm S}}{L_{\odot}} \right)
+\frac{1}{2}\, \log\left(\frac{M}{M_{\odot}} \right)
-\frac{3}{2}\,\log\left(\frac{D_{\rm ds}}{\rm kpc} \right).
\ee
In Table 1, we show $\log a- 0.4\, m_{\rm bol}$ for the listed sources
lensed by stars, with $M=M_{\odot}$ and $D_{\rm ds}=D_{\rm c}/2$.

Now we compute the volume fractions of space occupied by the
high-flux tubes, for the sources and lenses in Tables 1 and 2.
Let ${\cal F}_{\rm L}(f)$ and ${\cal F}_{\rm S}(f)$ denote the total
volume fractions of space in which the flux from the source exceeds $f$ 
due to gravitational lensing and due to being close to the source
respectively. A high-flux tube, in which the flux from the source exceeds
$f$, has a bolometric magnitude less than $m_{\rm bol}$. We find
\begin{eqnarray}
\log{\cal F}_{\rm L}&=& -8+ \log I(\alpha)+ 0.8  m_{\rm bol}
+2\log\left(\frac{L_{\rm S}}{L_{\odot}}\right)
+ \log\left(\frac{M}{M_{\odot}}\right) \nonumber\\
&& \hskip 0.5cm  +\log\left(\frac{D_{\rm c}}{1 {\rm kpc}}\right) 
+\log\left(\frac{n_{\rm S}}{1 {\rm pc}^{-3}}\right) 
+\log\left(\frac{n_{\rm L}}{1 {\rm pc}^{-3}}\right) \nonumber\\
\log\left(\frac{{\cal F}_{\rm L}}{{\cal F}_{\rm S}}\right)&=&
-8.86+ \log I(\alpha)+ 0.2  m_{\rm bol}
+\frac{1}{2}\log\left(\frac{L_{\rm S}}{L_{\odot}}\right)
+ \log\left(\frac{M_{\rm L}}{M_{\odot}}\right) \nonumber\\
&& \hskip 0.5cm  +\log\left(\frac{D_{\rm c}}{1 {\rm kpc}}\right) 
+\log\left(\frac{n_{\rm L}}{1 {\rm pc}^{-3}}\right), 
\end{eqnarray}
where $I(\alpha)$ is the reduction factor due to the finite source
size [see Eq.(\ref{eq:I(alpha)}) and Fig.5.].

In Fig.7(a), we plot $\log({\cal F}_{\rm L}/{\cal F}_{\rm S})$
for lensing by stars, for the same sources as in Fig.6
(with the same line types).
Fig.7(b) shows the corresponding $\log {\cal F}_{\rm L}$.
Fig.8(a) and (b) show $\log({\cal F}_{\rm L}/{\cal F}_{\rm S})$
and $\log({\cal F}_{\rm L})$ for lensing by giant black holes
for the same cosmological sources as in Fig.6 (with the same line types).
Only rough order of magnitude properties are used for the sources and lenses
in Figs.6-8. The largest effect comes from $\gamma$-ray bursts lensed by stars 
in our simple model (zero shear).

We note that although the high-flux volume fractions due to 
lensing are small, the corresponding absolute volumes can be large. 
Further, since materials move across the high flux tubes constantly, 
the fraction of material affected by EGLE is much higher than the static 
volume fractions. 
Finally, recall that we are neglecting shear, which might be 
important in some of the cases enumerated here.

\section{An example: destruction of dust grains in globular clusters}

As a specific example, we consider the
destruction of dust grains in the high-flux tubes produced by EGLE.
Let us consider a system of physical size $D_{\rm c}$,
with $N_{\rm S}$ bright sources, lensed by $N_{\rm L}$ lenses with mass $M$. 
The high-flux regions associated with being close to a source consist of 
$N_{\rm S}$ small spheres enclosing the sources, while the high-flux 
regions associated with EGLE consist of thin tubes stretching out 
from the $N_{\rm L}$ lenses distributed all over the system. 

For a given dust grain in this system,
the instantaneous probability that an EGLE produces flux greater than $f$ is given by
${\cal F}_{\rm L}(f)$, the volume fraction of space in tubes with flux
greater than $f$. The typical duration of this EGLE, $\Delta t$, is given by 
$2a/v$, where $a$ is the characteristic radius of the high-flux tubes, 
and $v$ is the typical velocity of the dust grain. The time between
such events is given by $t_f = \Delta t/{\cal F}_{\rm L}(f)$. 
For dust grains to be destroyed by EGLE, the following two conditions 
are sufficient: (1) One EGLE heats up a dust grain to sufficiently high 
temperature for a sufficient amount of time to destroy the dust grain;
(2) $t_f$ is less than the typical lifetime of a dust grain 
in the absence of EGLE. 

We now focus on the specific example of dust grain destruction within a
typical globular cluster.
In such a cluster, there are approximately 10 bright X-ray/UV
sources associated with accreting neutron stars, which are 
lensed by $10^6$ stars.  The bolometric luminosity of these sources is
thought to be of order their Eddington luminosities.
The parameter $\alpha$, which measures the maximum magnification of the 
source relative to the flux $f$ [see Eqs.(\ref{eq:alpha(f)}) and 
(\ref{eq:alphaAstro})], is given by
\begin{eqnarray}
\log(\alpha) &=& 21.38+0.8\,m_{\rm bol}(f)+ 2\,\log\left( 
\frac{L_{\rm S}}{10^5 L_{\odot}}\right)+ \log\left( \frac{M}
{M_{\odot}}\right)- 2\, \log\left( \frac{\rho}
{10^{-4} R_{\odot}}\right)
\nonumber\\
&&\hskip 2cm -3\, \log\left( \frac{D_{\rm c}}{1\,{\rm pc}}\right)
\end{eqnarray}
The optical depth for microlensing in this system is $\tau \sim
3\times 10^{-7}$, which means the shear $\gamma \sim 10^{-6}$
[see Eq.(\ref{eq:gamma})];
The dimensionless radius of the source is $R \sim 5 \times 10^{-6}$.
Clearly, shear is not important in this system [see Figs. 1 and 2].
Also, note that even though the lens, source, and target here are all
within a few parsecs of each other (separations very small compared
to most of the lensing situations currently discussed in the literature),
the Einstein ring radius of the lens is still much larger [by roughly
10 times] than the physical size of the star-lens; hence the
lens can be safely approximated as ``transparent''.
The formalism developed in the previous sections applies here.

The volume fraction of space occupied by high-flux tubes with
bolometric magnitude less than $m_{\rm bol}$ is
\begin{eqnarray}
\log{\cal F}_{\rm L}&=& 6.562+ \log I(\alpha)+ 0.8  m_{\rm bol}
+2\log\left(\frac{L_{\rm S}}{10^5 L_{\odot}}\right)
+ \log\left(\frac{M}{M_{\odot}}\right) \nonumber\\
&& \hskip 0.5cm  -5\, \log\left(\frac{D_{\rm c}}{1 {\rm pc}}\right) 
+\log\left(\frac{N_{\rm S}}{10}\right) 
+\log\left(\frac{N_{\rm L}}{10^6}\right) 
\end{eqnarray}
where $I(\alpha)$ is the reduction factor due to the finite source
size [see Eq.(\ref{eq:I(alpha)}) and Fig.5.].
The length of a high-flux tube is
\be
\log\left( \frac{D_{\rm d}^{\rm m}} {\rm pc} \right)=
\log\left( \frac{D_{\rm ds}}{\rm pc}\right) +\log g(\omega),
\ee
where $\omega=(D_{\rm ds}/D_{\rm c}) \alpha^{-1/3}$.
$g(\omega)$ is given by Eq.(\ref{eq:D_d(f)}) and shown in Fig.4. 
Note that if the high-flux tube is
longer than the size of the system $D_{\rm c}$, only the part of the
high-flux tube which lies inside the system contributes to
the dust grain heating in the system.
The characteristic radius $a$ of the high-flux tube is
\be
\log\left( \frac{a} {\rm cm} \right)= 16.88 +0.4\, m_{\rm bol}
+\log\left(\frac{L_{\rm S}}{10^5 L_{\odot}} \right)
+\frac{1}{2}\, \log\left(\frac{M}{M_{\odot}} \right)
-\frac{3}{2}\,\log\left(\frac{D_{\rm ds}}{\rm pc} \right).
\ee

Assuming 100\% efficiency in the absorption of photons by the dust
grains (as appropriate in the UV and soft X-ray bands), 
the relation of the grain temperature $T$ to the photon flux $f$
is given by (\cite{Draine84})
\be
f = 4 \langle Q(a_{\rm gr}, T) \rangle\, \sigma\, T^4,
\ee
where $\langle Q(a_{\rm gr}, T) \rangle$ is the Planck-averaged
emissivity for a dust grain with radius $a_{\rm gr}$, and $\sigma$ 
is the Stefan-Boltzmann constant.
Hence we have
\be
m_{\rm bol} = -10 \, \log\left( \frac{T\, \langle Q(a_{\rm gr}, T) 
\rangle ^{1/4}}{0.56 {\rm K}}\right).
\ee
For a silicate grain,
$\langle Q(a_{\rm gr}, T) \rangle \la 0.4a_{\rm gr}/\mu$m 
for $ 100 {\rm K}<T<1000{\rm K}$;
we use this upper limit on the emissivity below.

For a dust grain moving with velocity $v$, the duration of one EGLE 
is $\Delta t =2 a/v$. We find
\be
\frac{\Delta t}{\rm sec} = 260.5 \left( \frac{T}{300 {\rm K}}\right)^{-4}
\left( \frac{a_{\rm gr}}{0.1 \mu{\rm m}}\right)^{-1}
\left( \frac{L_{\rm S}}{10^5 L_{\odot}}\right)
\left( \frac{M}{M_{\odot}}\right)^{1/2}
\left( \frac{D_{\rm ds}}{0.5 {\rm pc}}\right)^{-3/2}
\left( \frac{v}{ 5 {\rm km}/{\rm sec}}\right)^{-1}.
\ee
In Fig.9, we plot $t_f = \Delta t/{\cal F}_{\rm L}(f)$,
the mean time between EGLEs, as function of the resulting dust 
grain temperature for three typical values of the grain radius.
The bend at low temperature is due to the cut-off in length of the high-flux 
tubes longer than $D_{\rm c}=1$pc (the lower the flux or temperature, 
the longer the corresponding EGLE tube).
A silicate grain with $a_{\rm gr}$ of order 0.1 $\mu$m can be heated to
over 300$\,$K for a few minutes, once every $5\times 10^7$ years;
this would be sufficient to destroy any icy components of the grain. 
Smaller silicate grains of, say, $0.01 \mu$m radius would be more fragile 
and would
be heated to over 1000$\,$K for 21 seconds once every $8.5\times 10^8$
years, perhaps sufficient to destroy them.  
Thus, EGLE may provide a mechanism for explaining the observed absence of 
dust in globular clusters (Knapp, Gunn and Connolly 1995).

Note that although we used the maximum cross-section of the high-flux
tubes in estimating the EGLE durations, we also assumed that a
dust grain moves through the high-flux tube exactly tangentially; 
if the dust grain enters the high-flux tube at an angle, the duration 
of the event is longer. In addition, we used the $minimum$ flux $f$ and 
temperature $T$ which define the event
in our calculations; the peak values during each event will be
substantially higher than those at the edge of the tube.
Furthermore, the dust grain emissivity is likely lower than the extreme
value we
adopted. Therefore, our estimate of the EGLE effects on grains in globular
clusters
is rather conservative. 

We have also ignored the interaction of globular cluster EGLE with the
gaseous intracluster medium that would presumably have accompanied any
putative dust component.  Of course, EGLE conserve photons and thus
total ionizing events (for example), so we do not expect such dramatic
effects as when there are threshold phenomena such as dust grain evaporation,
but the spatial redistribution of the photons might still have some
subtle consequences.
A more accurate and detailed examination of this and other possible
EGLE effects is beyond the scope of the present paper.

We thank Bruce Draine for very instructive discussions of possible dust
grain effects.
We also thank the ApJ Scientific Editor E. L. Wright and the anonymous 
referee for comments which led to a clearer presentation of our results.
Y.W. is supported by the DOE and NASA under Grant NAG5-2788.
E.L.T. gratefully acknowledges support from NSF grant AST94-19400.
 
\newpage
\begin{table}[h]
\caption{List of possible EGLE sources. $\log\alpha$ and $\log a$
are computed with $M=M_{\odot}$ and $D_{\rm ds}=D_{\rm c}/2$.}
\renewcommand
\arraystretch{1.25}
\begin{center}
\begin{tabular}{|r||c|c|c|c|c|c|}
\hline
  & $\rho/R_{\odot}$ & $L_{\rm S}/L_{\odot}$ & $D_{\rm c}/$kpc & 
$n_{\rm S}/\mbox{pc}^{-3}$ & 
$\begin{array}{ll}\log\alpha \\-0.8m_{\rm bol}\end{array}$ &
$\begin{array}{ll}\log(a/{\rm cm})\\-0.4m_{\rm bol} \end{array}$ \\ \hline
QSO $\left\{\begin{array}{ll} \mbox{X-ray} \\\mbox{UV-opt} \end{array}\right.$ 
& $\begin{array}{ll} 10^{2}\\ 10^{4} \end{array} $
&$\begin{array}{ll} 10^{12} \\10^{13} \end{array}$ & $10^6$ 
&$3\times 10^{-22}$
&$\begin{array}{ll} -3.62\\ -5.62 \end{array} $ 
&$\begin{array}{ll} 10.83 \\11.83 \end{array} $		\\ \hline
$\gamma$-ray bursts & $10^{-5}$-$0.1$ & $10^{18\mbox{-}20}$	 
	    & $10^6$   & $10^{-31}$ &
14.38-26.38 &16.83-18.83	\\ \hline
Galactic supernovae & 	10$^3$ & 	$10^{10}$ 
	    & 10   	&  $10^{-14}$ &
5.38 &	16.33	\\ \hline
neutron stars $\left\{ \begin{array}{ll} \mbox{ X-ray} \\
				\mbox{ radio}
				\end{array} \right.$
 & $10^{-5}$ & $\begin{array}{ll}10^5  \\10^6  \end{array}$
	    & $10^{-2}$-1   &  $10^{-9}$ &
$\begin{array}{ll}\mbox{14.38-20.38}\\ \mbox{16.38-22.38}
\end{array}$ & $\begin{array}{ll} \mbox{12.83-15.83}\\
\mbox{13.83-16.83}\end{array}$		\\ \hline
hot stars $\left\{\begin{array}{ll}\mbox{O} \\ \mbox{B} \end{array}\right.$
& $\begin{array}{ll} 10 \\ 4 \end{array}$
& $\begin{array}{ll} 10^5 \\10^3 \end{array}$
 & 1  &  $\begin{array}{ll} 10^{-8}  \\10^{-4} \end{array}$  &
$\begin{array}{ll} 2.38 \\-0.824 \end{array}$ &	
$\begin{array}{ll} 12.83 \\ 10.83 \end{array}$	\\ \hline
\end{tabular}
\tablecomments{$D_{\rm c}$ is the size of the volume in which the sources,
the lenses, and the targets are all uniformly distributed.}
\end{center}

\end{table}

\begin{table}[h]
\caption{List of possible EGLE lenses}
\begin{center}
\begin{tabular}{|c||c|c|}
\hline
  & $M$ & $n_{\rm L}$ \\ \hline
$\begin{array}{ll}\mbox{giant black holes} \\ \mbox{(cosmological)}
\end{array}$ & $10^{6-8} M_\odot$ &  $10^3$/(Mpc)$^3$ \\ \hline
stars $\begin{array}{ll}
\mbox{(galactic average)} \\
 \mbox{(cosmological average)}\end{array}$
& 0.1-1 $M_\odot$ &  
$\begin{array}{ll}\mbox{1.4 /(pc})^{3}\\
10^{11} \mbox{/(Mpc)}^{3}\end{array}
$\\ \hline
\end{tabular}
\end{center}
\end{table}

%

\clearpage

\figcaption[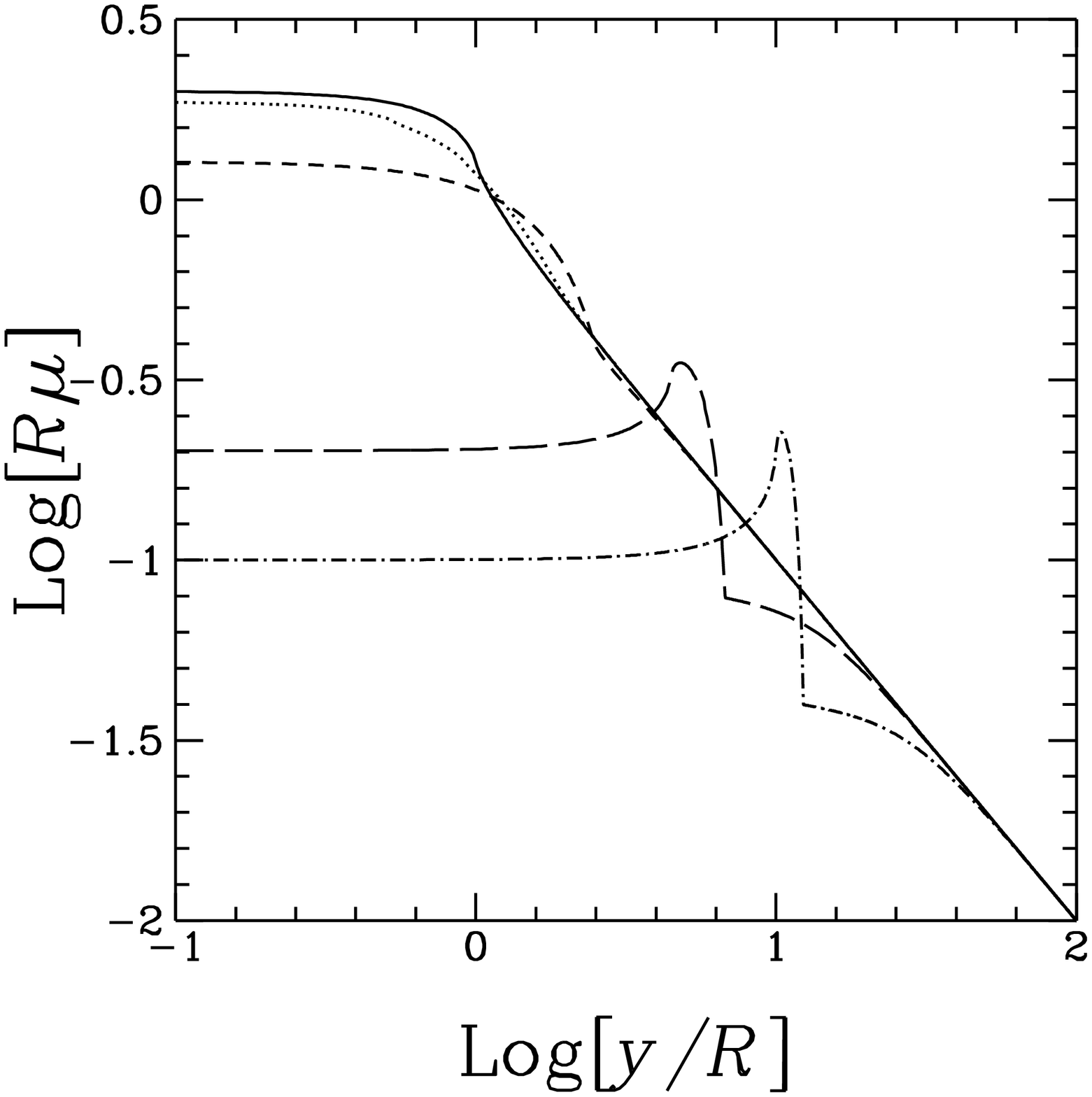]{Typical lightcurves for shear $\gamma=0$ (solid line), 
$\gamma=R/2$ (dotted line), $\gamma=R$ (short dashed line), $\gamma=5R$ 
(long dashed line), and $\gamma=10R$ (dot-dashed line). The dimensionless
source size $R\ll 1$. $y$ and $\mu$ are the angular position and the 
maginification of the source respectively.}

\figcaption{Cross-sections of magnification for lightcurves in Fig.1
(with the same line types).}

\figcaption{Cross-section $\sigma(f, \rho, D_{\rm d})$ with $\alpha(f)=4$, for 
$D_{\rm ds}=0.2\,D_{\rm c}$ (solid line), $0.5\,D_{\rm c}$ (long dashed line).}

\figcaption{The scaled length of the high-flux ($>f$) tube,
$g(\omega)= D_{\rm d}^{\rm m}(f)/D_{\rm ds}$, as a function of
$\omega = \left(D_{\rm ds}/D_{\rm c}\right) \,\alpha^{-1/3}(f)$. }

\figcaption{$I(\alpha)= V_{\rm L}(f, \rho)/V_{\rm L}(f, \rho=0)$, the
reduction factor in EGLE volumes due to finite source size. }

\figcaption{The high-flux ($>f$) tube's length $D_{\rm d}^{\rm m}$ 
versus $m_{\rm bol}(f)$ for various sources lensed by stars.
The sources are $\gamma$-ray bursts (solid line),
QSO (X-ray) (dotted line), QSO (UV-opt) (short dashed line), 
Galactic supernovae (long dashed line), neutron stars (dot-short dashed line),
hot O stars (dot-long dashed line), and hot B stars 
(short dash-long dashed line) respectively.}

\figcaption{Lensing by stars of the same sources as in Fig.6 (with the
same line types). (a) $\log({\cal F}_{\rm L}/{\cal F}_{\rm S})$.
(b) $\log {\cal F}_{\rm L}$.}

\figcaption{Lensing by giant black holes of the same cosmological
	sources as in Fig.6 (with the same line types). 
	(a) $\log({\cal F}_{\rm L}/{\cal F}_{\rm S})$.
(b) $\log({\cal F}_{\rm L})$.}

\figcaption{The time between two EGLEs as function of the silicate grain temperature, for X-ray/UV sources lensed by stars in a typical globular
cluster. The values of the grain radius are labeled on the curves. }



\clearpage

\setcounter{figure}{0}
\plotone{fig1.eps}
\figcaption[fig1.eps]{Typical lightcurves for shear $\gamma=0$ (solid line), 
$\gamma=R/2$ (dotted line), $\gamma=R$ (short dashed line), $\gamma=5R$ 
(long dashed line), and $\gamma=10R$ (dot-dashed line). The dimensionless
source size $R\ll 1$. $y$ and $\mu$ are the angular position and the 
maginification of the source respectively.}

\clearpage

\plotone{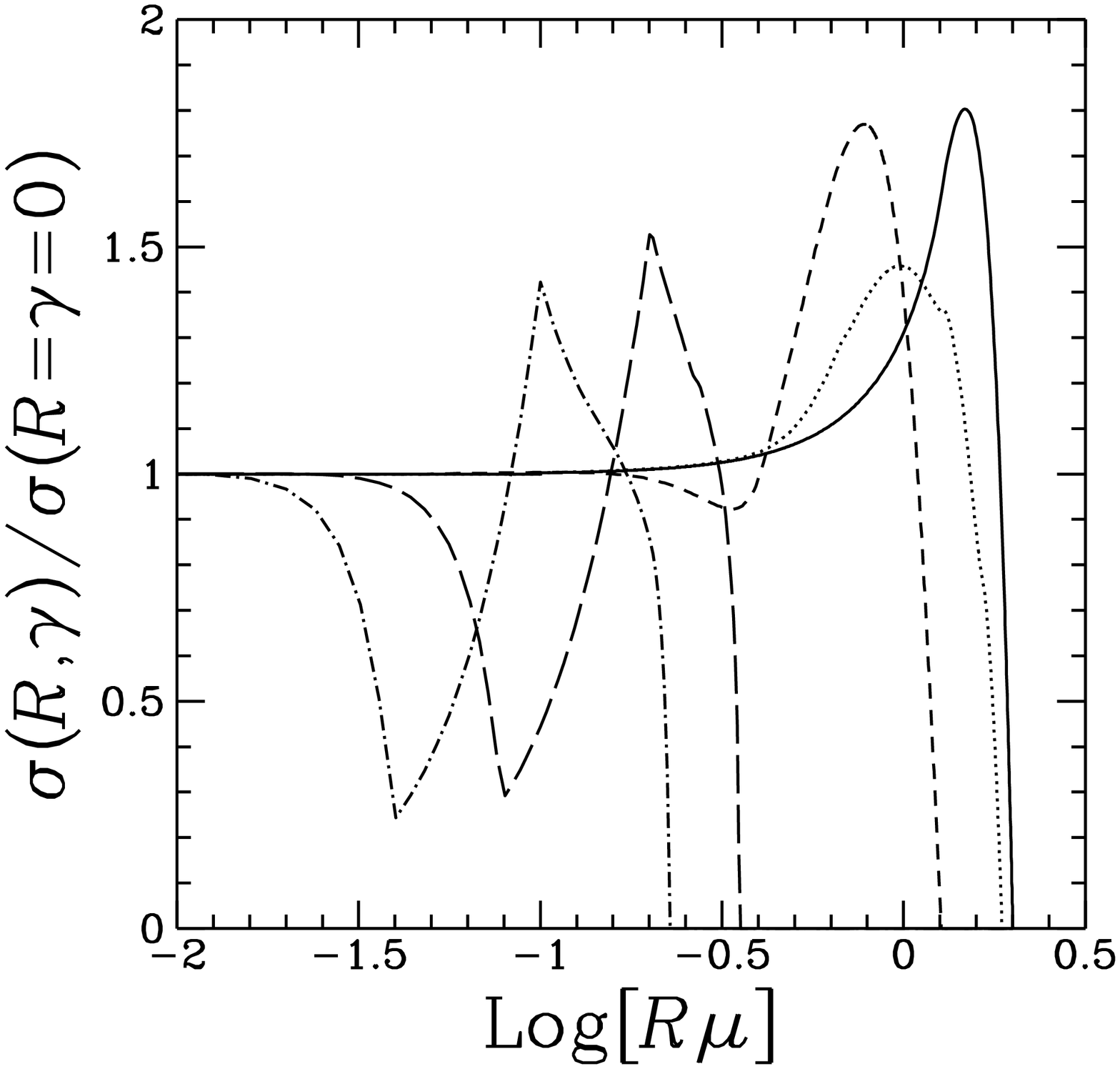}
\figcaption{Cross-sections of magnification for lightcurves in Fig.1
(with the same line types).}

\plotone{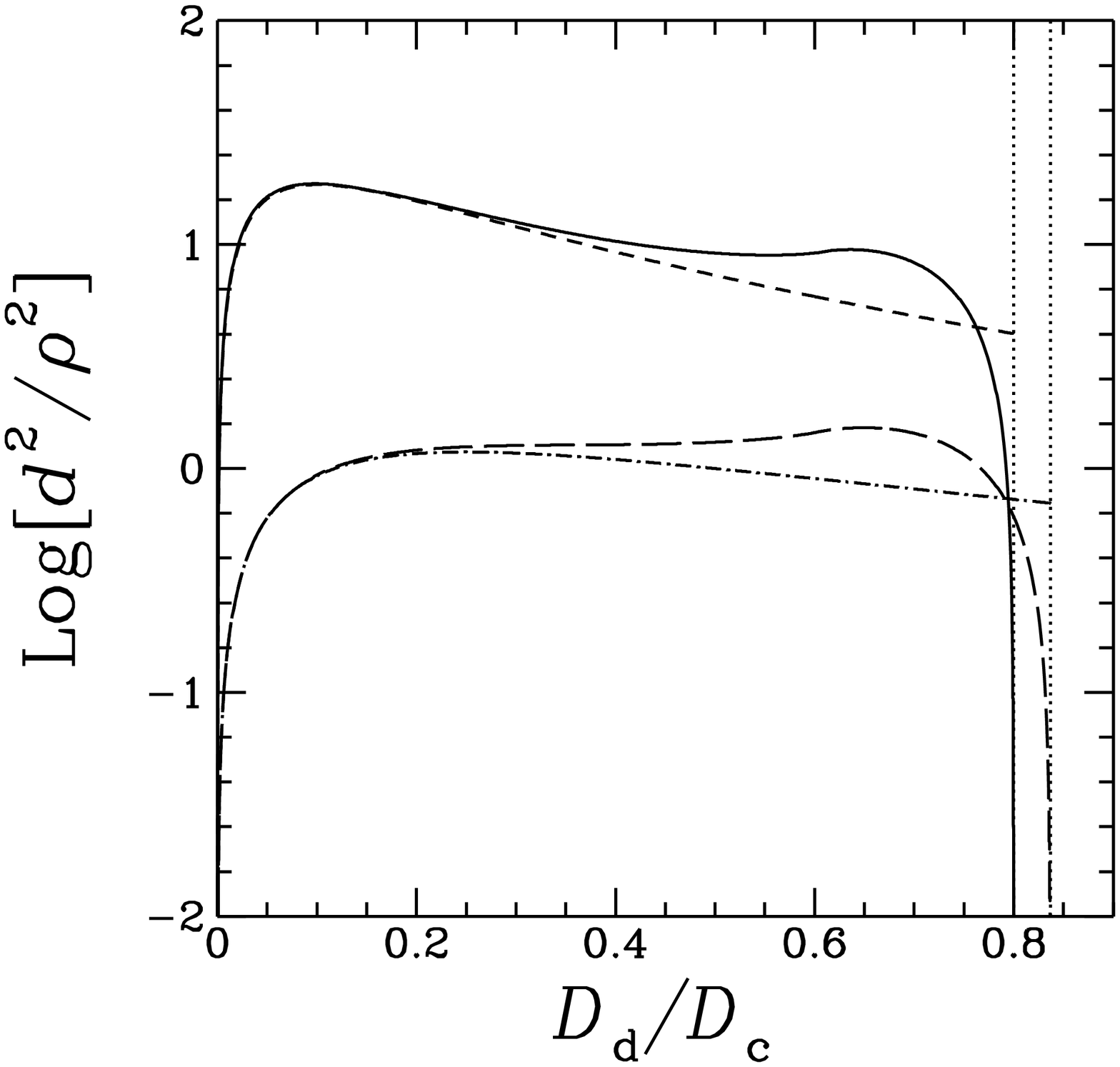}
\figcaption{Cross-section $\sigma(f, \rho, D_{\rm d})$ with $\alpha(f)=4$, for 
$D_{\rm ds}=0.2\,D_{\rm c}$ (solid line), $0.5\,D_{\rm c}$ (long dashed line).}

\plotone{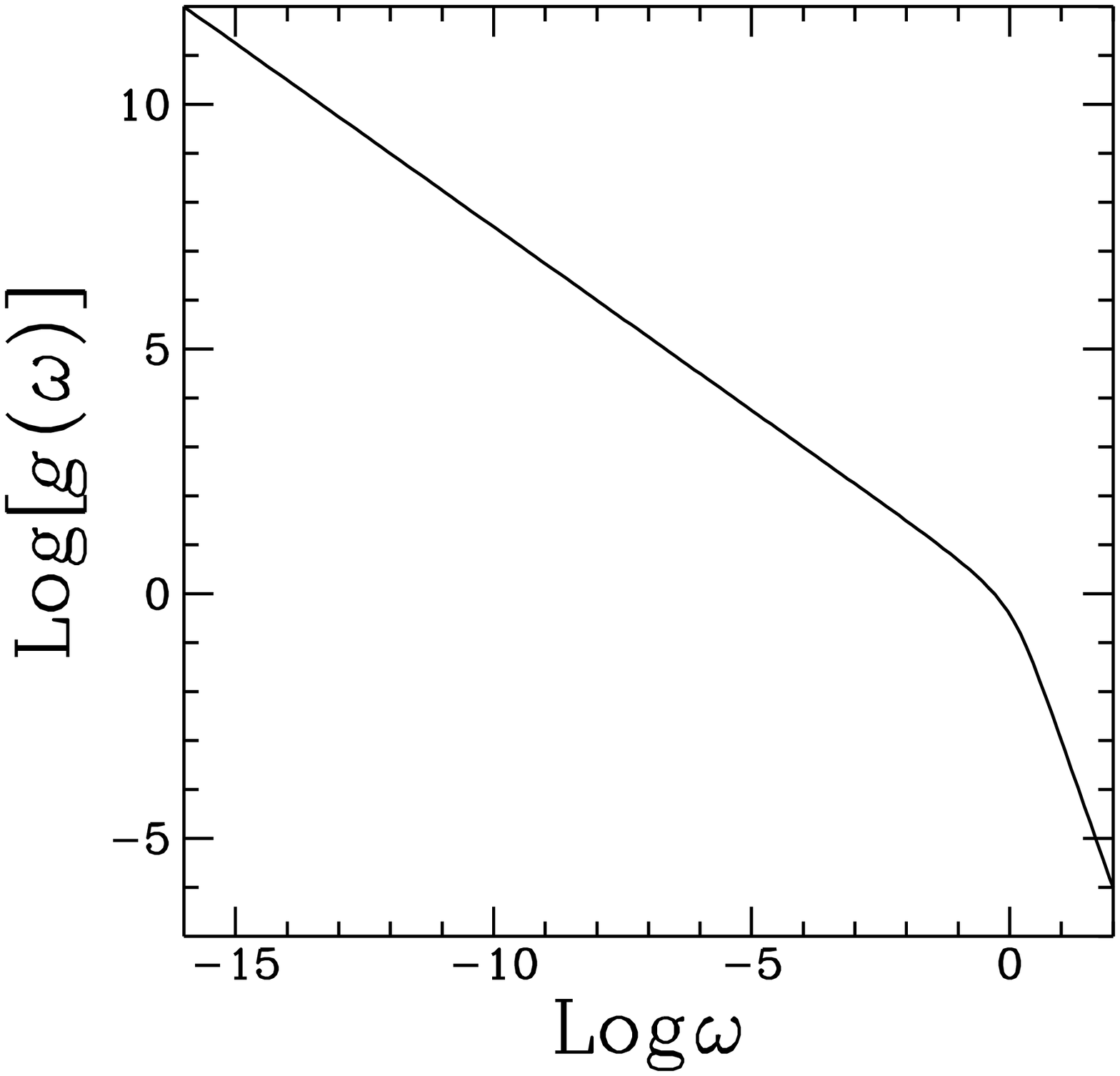}
\figcaption{The scaled length of the high-flux ($>f$) tube,
$g(\omega)= D_{\rm d}^{\rm m}(f)/D_{\rm ds}$, as a function of
$\omega = \left(D_{\rm ds}/D_{\rm c}\right) \,\alpha^{-1/3}(f)$. }

\plotone{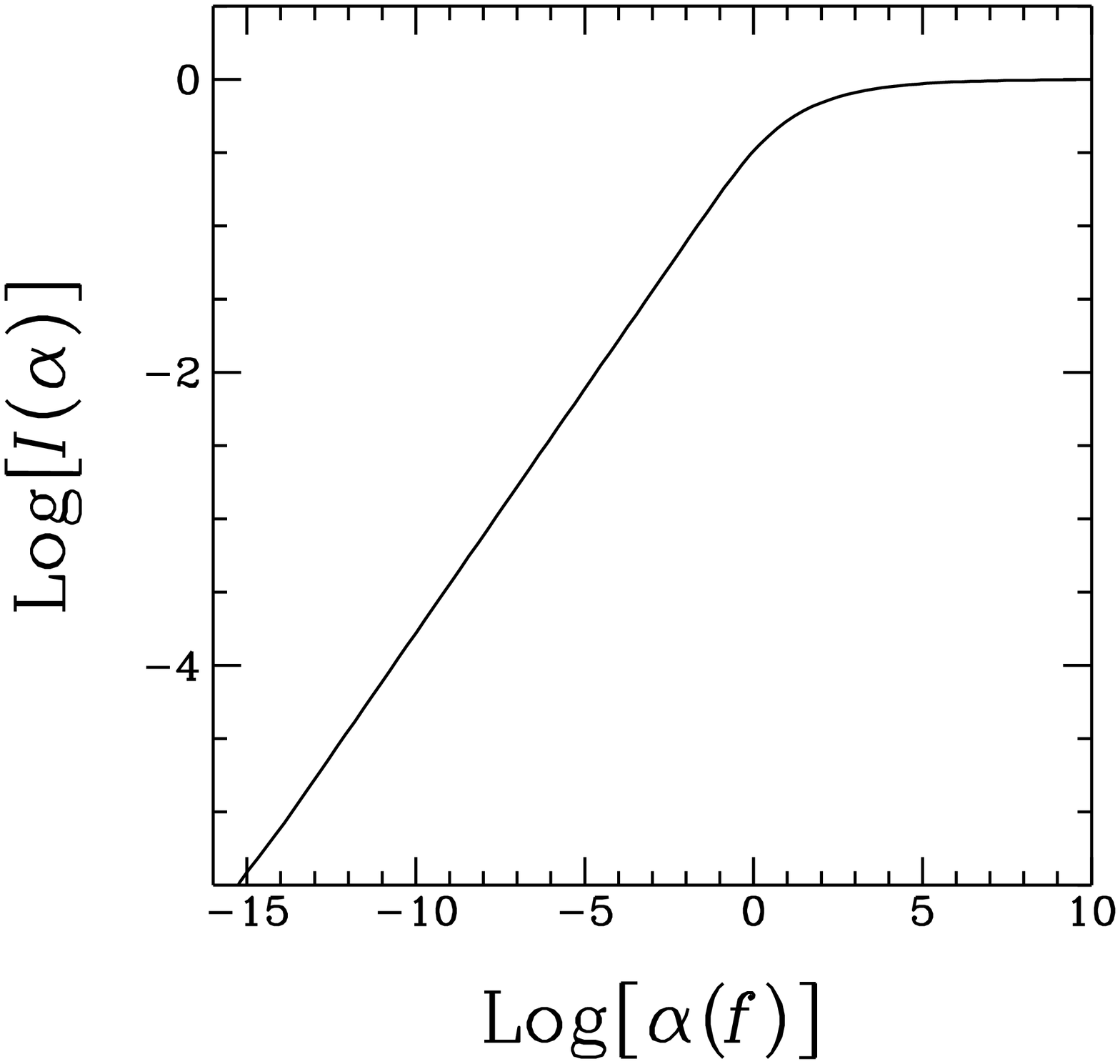}
\figcaption{$I(\alpha)= V_{\rm L}(f, \rho)/V_{\rm L}(f, \rho=0)$, the
reduction factor in EGLE volumes due to finite source size. }

\plotone{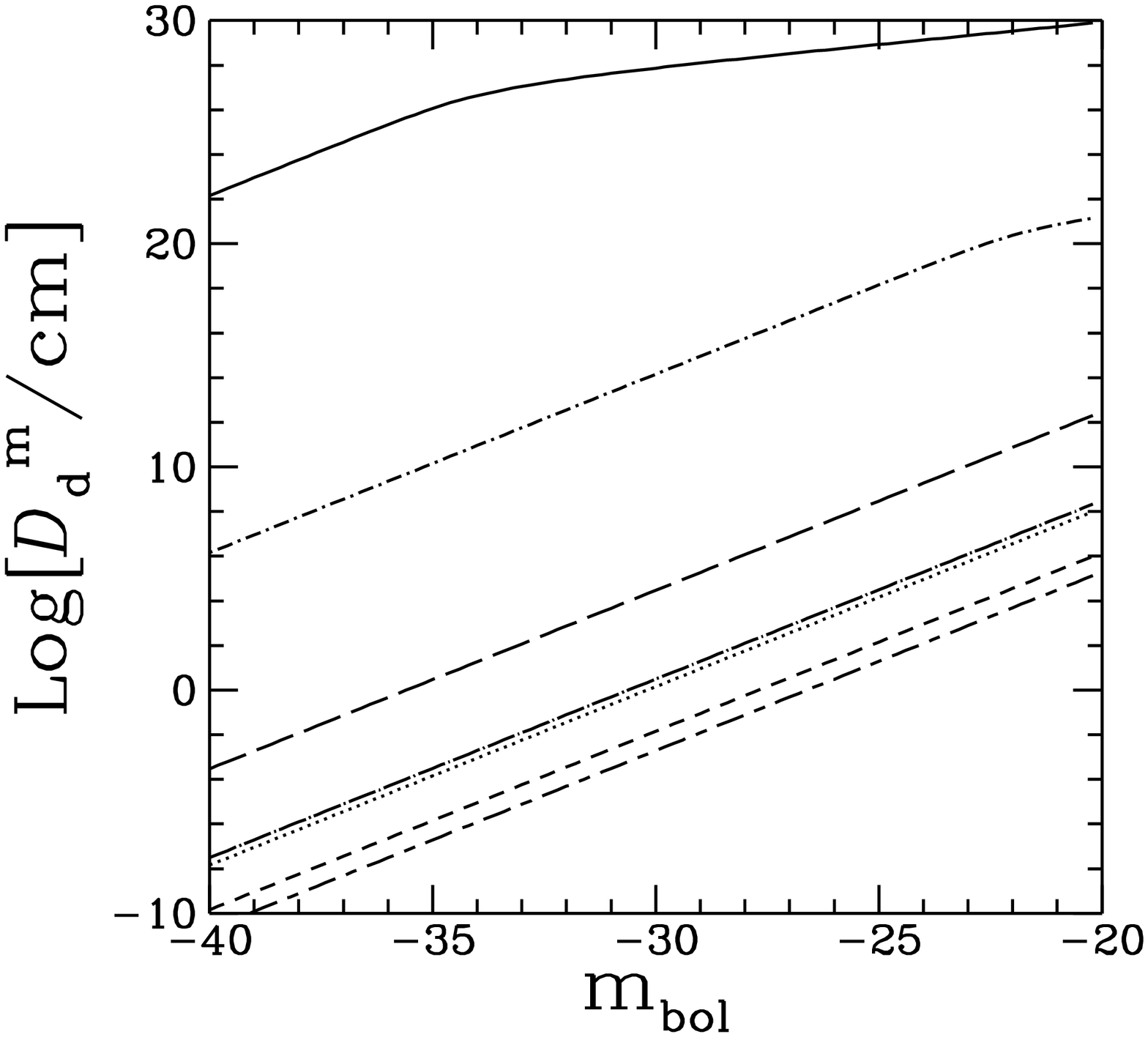}
\figcaption{The high-flux ($>f$) tube's length $D_{\rm d}^{\rm m}$ 
versus $m_{\rm bol}(f)$ for various sources lensed by stars.
The sources are $\gamma$-ray bursts (solid line),
QSO (X-ray) (dotted line), QSO (UV-opt) (short dashed line), 
Galactic supernovae (long dashed line), neutron stars (dot-short dashed line),
hot O stars (dot-long dashed line), and hot B stars 
(short dash-long dashed line) respectively.}

\plotone{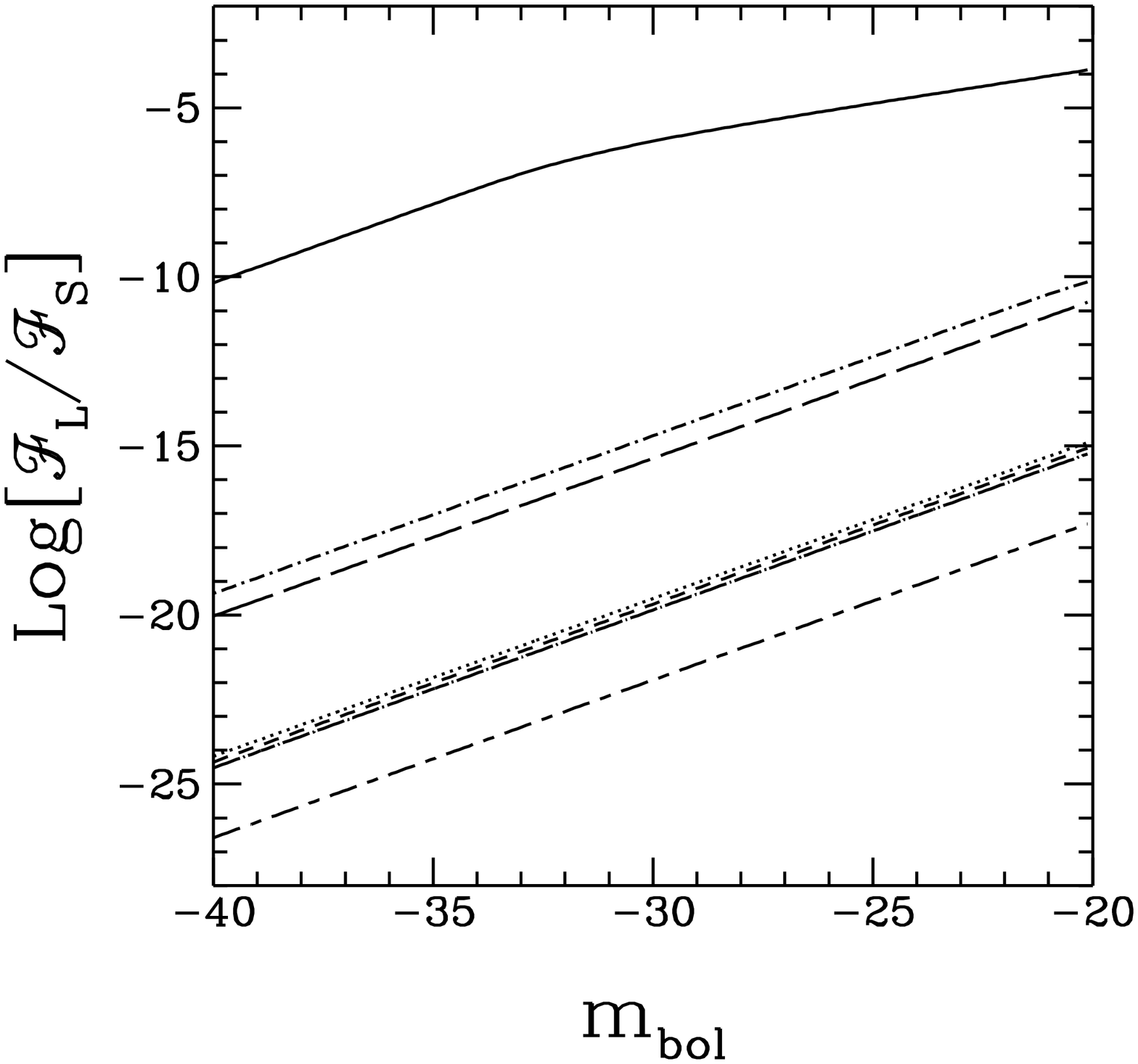}
\figcaption{Lensing by stars of the same sources as in Fig.6 (with the
same line types). (a) $\log({\cal F}_{\rm L}/{\cal F}_{\rm S})$.}

\setcounter{figure}{6}
\plotone{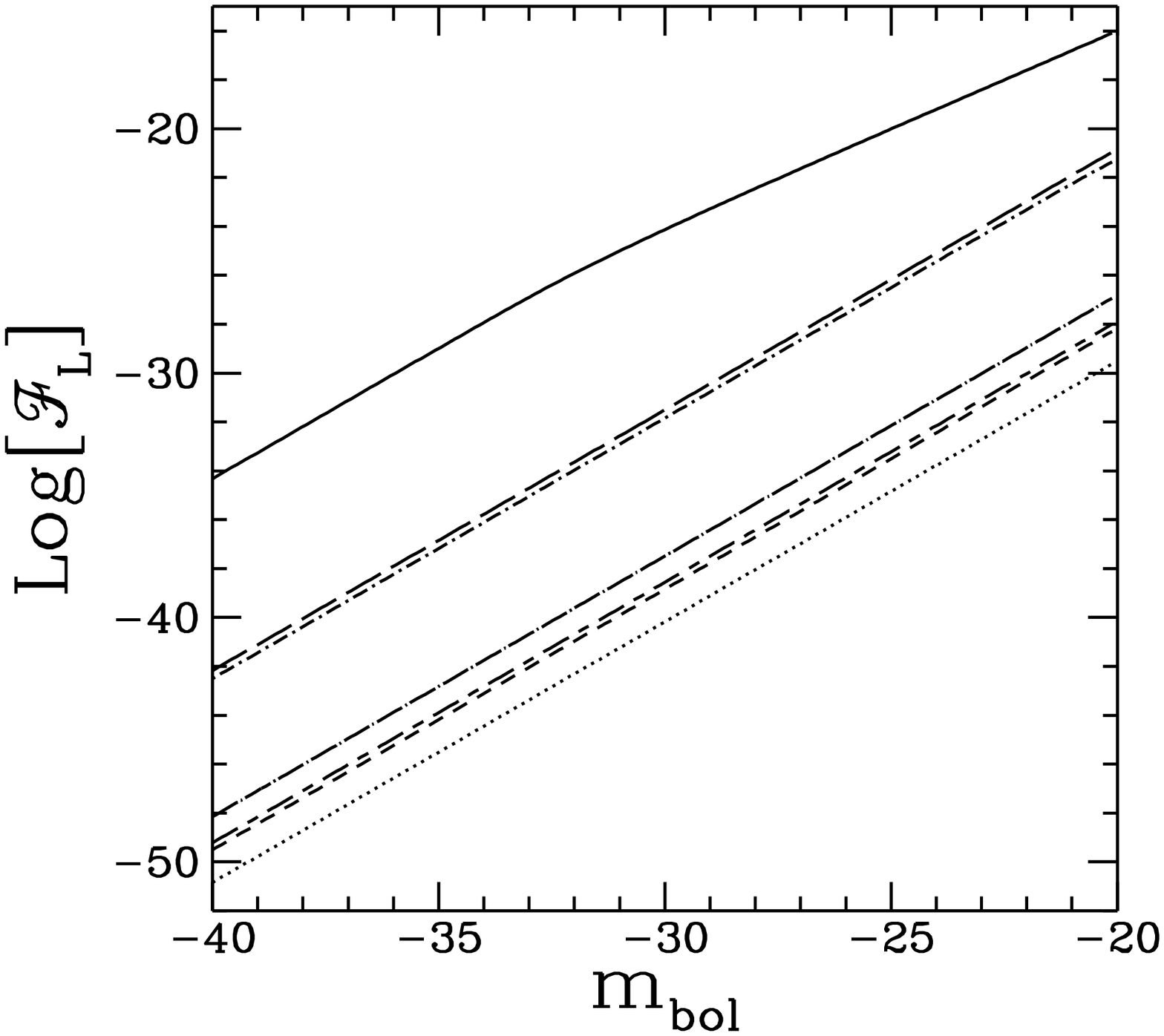}
\figcaption{(b) $\log {\cal F}_{\rm L}$.}

\plotone{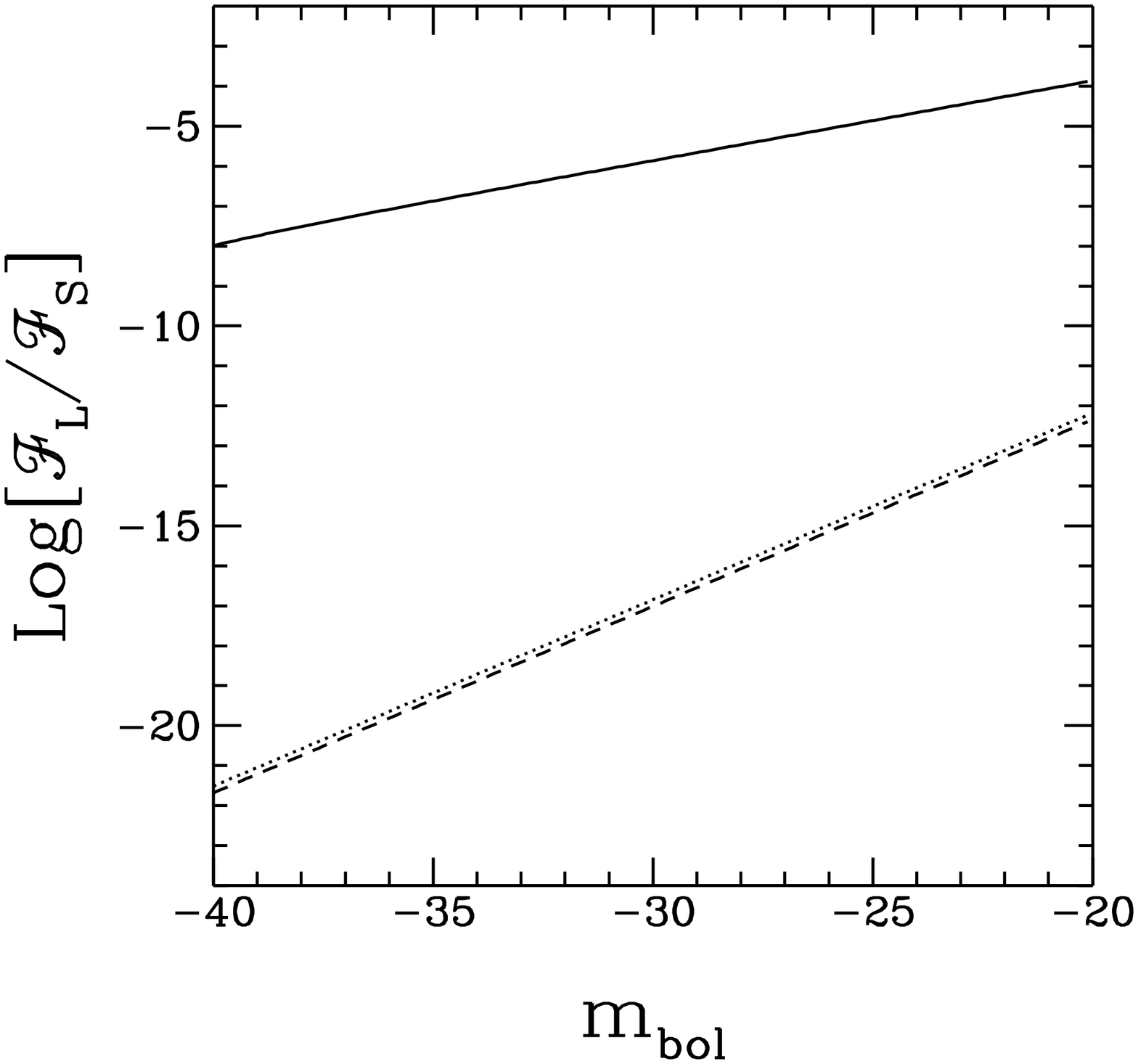}
\figcaption{Lensing by giant black holes of the same cosmological
	sources as in Fig.6 (with the same line types). 
(a) $\log({\cal F}_{\rm L}/{\cal F}_{\rm S})$.}
\setcounter{figure}{7}

\plotone{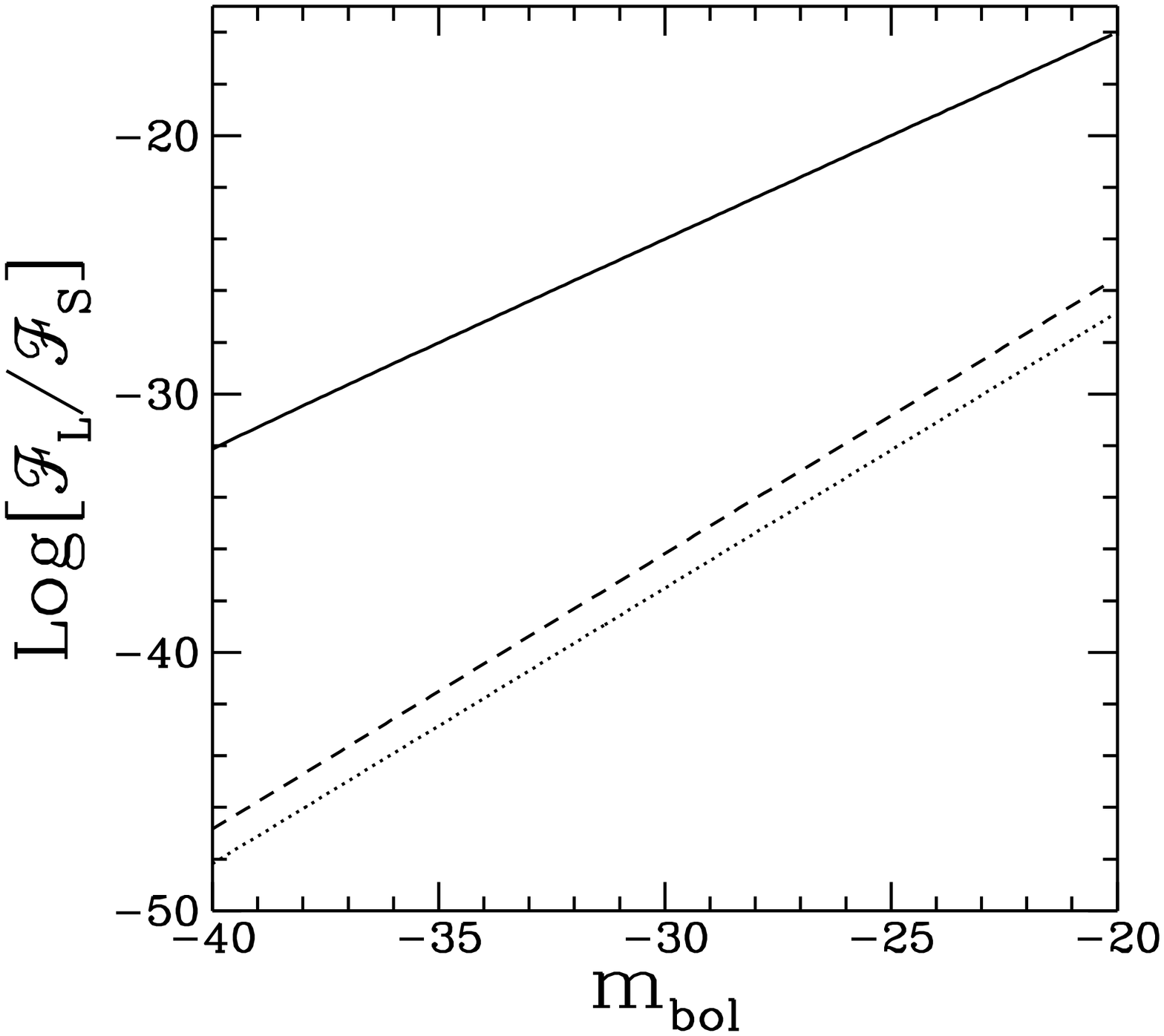}
\figcaption{(b) $\log({\cal F}_{\rm L})$.}

\plotone{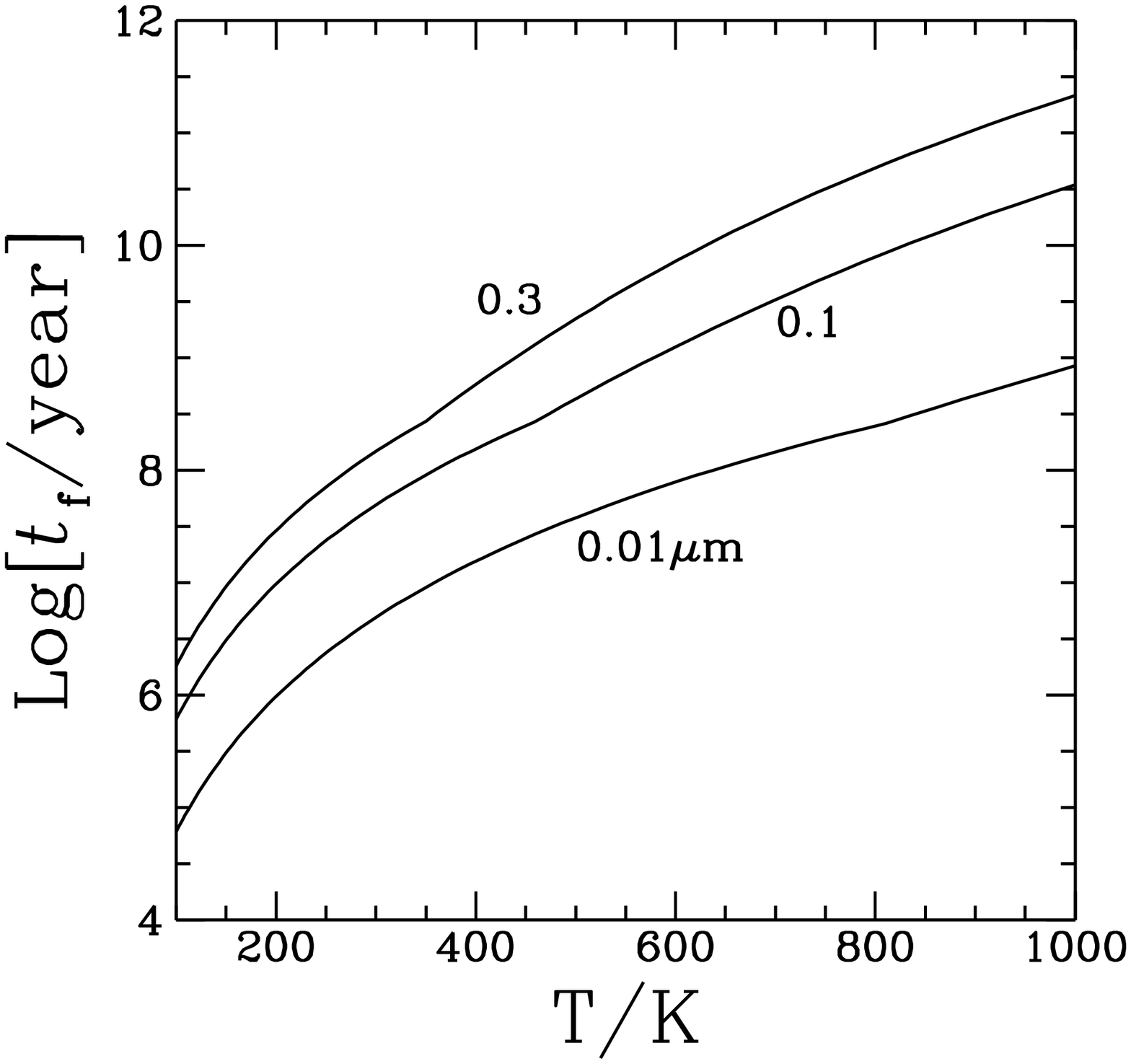}
\figcaption{The time between two EGLEs as function of the silicate grain temperature, for X-ray/UV sources lensed by stars in a typical globular
cluster. The values of the grain radius are labeled on the curves. }


\clearpage


\begin{thebibliography}{}

\bibitem[Blandford and Narayan 1992]{BNrev92}
Blandford, R.D. and Narayan, R. (1992), Ann. Rev. Astron. Astrophys., 30, 311.
 
\bibitem[Bontz 1979]{BO79.1}
Bontz, R.J. (1979), \apj, 233, 402.


\bibitem[Chang and Refsdal 1979]{CH79.1}
 Chang, K. and Refsdal, S. (1979), Nature, 282, 561.


\bibitem[Chang 1984]{CH84.1}
 Chang, K (1984), Astr. Ap., 130, 157.

 
\bibitem[Chang and Refsdal 1984]{CH84.2}
 Chang, K. and Refsdal, S. (1984), Astr. Ap., 132, 168.


\bibitem[Draine and Lee 1984]{Draine84}
Draine, B.T. and Lee, H.M. (1984), \apj, 285, 89.

\bibitem[Knapp, Gunn and Connolly 1995]{Knapp95}
Knapp, G.R., Gunn, J.E. and Connolly, A.J. (1995), preprint POP-614.

\bibitem[Mao and Paczynski 1992]{gamma-rayBurst}
Mao, S. and Paczynski, B. (1992), \apj, 388, L45.

\bibitem[Schneider and Weiss 1987]{SC87.3}
 Schneider, P. and Weiss, A. (1987), Astr. Ap., 171, 49.


\bibitem[Schneider 1987]{SC87.4}
 Schneider, P. (1987), Astr. Ap., 179, 71.


\bibitem[Schneider et al.\ 1992]{Book92}
Schneider, P., Ehlers, J., Falco, E.E., (ed. Springer-Verlag, Berlin, 1992)\/,  ``Gravitational lenses''.

 
\bibitem[Subramanian and Chitre 1985]{SU85.1} 
Subramanian, K. and Chitre, S.M. (1985), \apj, 289, 37.

\bibitem[Turner 1980]{Turner80}
Turner, E. L. (1980), \apj, 242, L135.
\bibitem[Turner et al.\ 1984]{Turner84}
Turner, E. L., Ostriker, J. P., and Gott, J. R. (1984), \apj, 284, 1.
\bibitem[Wang and Turner, in preparation]{Turner96}
Wang, Y. and Turner, E. L., in preparation.
\end{thebibliography}
\end{document}